# Potassium under pressure: electronic origin of complex structures

V F Degtyareva[1] and O Degtyareva[2]

[1] Institute of Solid State Physics, Russian Academy of Sciences, Chernogolovka, 142432 Russia
[2] CSEC and School of Physics, University of Edinburgh, The King's Buildings, Mayfield Road, Edinburgh EH9 3JZ, UK

E-mail: degtyar@issp.ac.ru

**Abstract**

Recent high-pressure x-ray diffraction studies of alkali metals revealed unusual complex structures that follow the body-centered and face-centered cubic structures on compression. The structural sequence of potassium under compression to 1 Megabar is as follows:
   bcc - fcc - h-g ($tI19^*$), $hP4$ - $oP8$ - $tI4$ - $oC16$.
We consider configurations of Brillouin-Jones zones and the Fermi surface within a nearly-free-electron model in order to analyze the importance of these configurations for the crystal structure energy that contains two main contributions: electrostatic (Ewald) and electronic (band structure) energies. The latter can be lowered due to a formation of Brillouin zone planes close to the Fermi surface opening an energy gap at these planes. Under pressure, the band structure energy term becomes more important leading to a formation of complex low-symmetry structures. The stability of the *post-fcc* phases in K is attributed to the changes in the valence electron configuration implying the overlap of valence band with the upper core electrons. This effect offers an understanding of structural complexity of alkali metals under strong compression.

## 1. Introduction

As recent experiments show, structure and properties of material change dramatically under pressure (see review papers [1-3] and references therein). Periodic table of elements if considered in compressed state looks very different from what we are familiar with at ambient conditions. The elements on the right side of the Periodic table become close-packed metals and superconductors, like Si and Ge, whereas simple metallic structures such as bcc and fcc of the elements from the left side of the Periodic table (groups 1-2) transform under pressure to open and complex structures. Na, one the best metal in terms of conductivity and reflectivity, becomes insulating and transparent at around 2 Mbar [4]. Lithium becomes superconducting [5] below 20K at pressures around 0.5 Mbar and at higher pressures shows a series of complex structures even with a semiconducting behaviour [6,7].

Differences between the atomic volumes of elements across one period level out (Figure 1). This means that the elements that have the large volumes at ambient conditions have much higher compressibility. K at 112 GPa, the highest pressure reached for this element experimentally, is compressed by 6.7 times and occupies 0.15 of its initial atomic volume – the largest compression among elements achieved so far (Figure 2). For the monovalent post-transition element Cu the value of $V/V_0$ at this pressure is only 0.7. Such large compression of K is accompanied by the changes in electronic structure where the electronic levels overlap. Some suggestions of the mechanism have been made, however they don't explain all the observations and questions remain. Interestingly, similar complex structural arrangements are observed for the elements from the right side of the periodic table (groups 13-16), with atomic volumes close to the values measured for the group 1 and 2 elements (Figure 1).

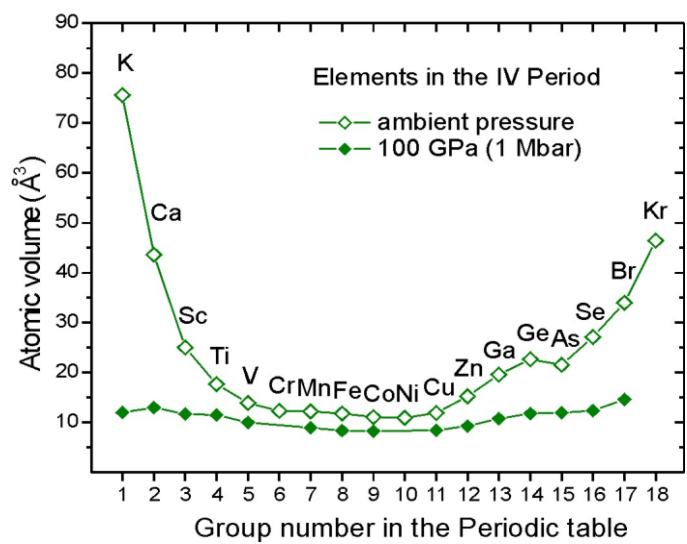

Figure 1. Atomic volumes of the elements as a function of the group number for the elements in the IV period of the Periodic Table. Open symbols correspond to the volume at ambient conditions. Filled symbols correspond to the pressure of 100 GPa (1Mbar) and are taken from the experimental measurements done by authors and found in the literature (see review papers [1-3] and references therein).

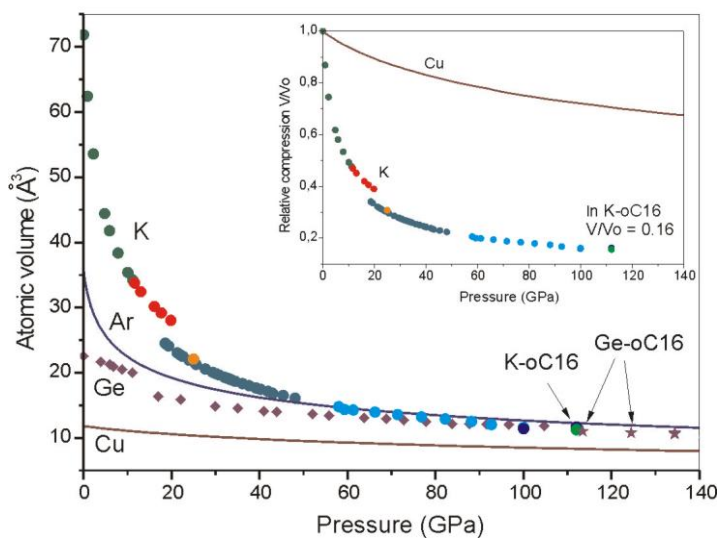

Figure 2. Atomic volumes of potassium on compression in comparison with Ar, Ge and Cu (K: [8-10], Cu: [11], Ar: [12], Ge: [13,14]). Relative compressibility of K and Cu is shown in the insert.

Argon is the closed-shell element next to potassium in the Periodic Table. Experimental P-V data for argon [12] are also shown in Figure 2. At 20 GPa the atomic volume of K (fcc, around the II-III transition) is ~45% *larger* than that of Ar, and above ~60 GPa the atomic volume of K becomes *smaller* than that of Ar, and on further compression to 112 GPa is ~7% smaller (in the K-$oC$16 phase). Volume changes with pressure for K are compared with P-V data for Ge to underline the similarity of their atomic volumes at pressure of ~100 GPa when they both adopt the same crystal structure $oC$16 where the compression of Ge is $V/V_0 \sim 0.5$ [13,14]. These volume changes in compressed K imply significant changes in electronic structure.

Potassium belongs to the group-I elements – alkali metals – lying in the intermediate position between light (Li, Na) and heavy (Rb, Cs) alkalis. At ambient condition all alkali elements crystallize in the body-centered cubic structure (bcc) and considered as free-electron metals. The first observed transformation on compression for all alkalis was to face-centered cubic structure (fcc) and further transitions were to complex structures summarized in Table 1. In addition to phases in Table 1, several very complex structures were observed for sodium at pressures above 100 GPa in the vicinity of the minimum of the melting curve determined as $tI$50, $oC$120, $mP$512, $aP$90 [15].

**Table 1.** Sequences of structural transformations on pressure increase for group-I elements

| | |
|---|---|
| **Li** | $\quad$ 7.5 $\quad\quad$ 39 $\quad\quad$ 42 $\quad\quad$ 60 $\quad\quad$ 70 $\quad\quad$ 95<br>bcc → fcc → $hR$1 → $cI$16 ➤ $oC$88 → $oC$40 → $oC$24 < 125 |
| **Na** | $\quad$ 65 $\quad\quad$ 104 $\quad\quad$ 117 $\quad\quad$ 125 $\quad\quad$ 180<br>bcc → fcc → $cI$16 ➤ $oP$8 → h-g ($tI$19*) → $hP$4 < 200 GPa |
| **K** | $\quad$ 11.6 $\quad\quad$ 20 $\quad\quad\quad$ 54 $\quad\quad$ 90 $\quad\quad$ 96<br>bcc → fcc ➤ h-g ($tI$19*) → $oP$8 → $tI$4 → $oC$16 < 112 GPa<br>$\quad\quad\quad\quad$ 25 $\quad\quad$ 35<br>$\quad\quad\quad\quad$ ➤ $hP$4 → |
| **Rb** | $\quad$ 7 $\quad\quad$ 13 $\quad\quad$ 17 $\quad\quad$ 20 $\quad\quad$ 48<br>bcc → fcc → $oC$52 ➤ h-g ($tI$19*) → $tI$4 → $oC$16 < 70 GPa |
| **Cs** | $\quad$ 2.4 $\quad\quad$ 4.2 $\quad\quad$ 4.3 $\quad\quad$ 12 $\quad\quad$ 72<br>bcc → fcc → $oC$84 ➤ $tI$4 → $oC$16 → dhcp < 223 GPa |

Notes: For experimental data see review papers [1-3] and references therein, data on Li are taken from [6]. The numbers above the arrows show the transition pressures in GPa. The crystal structures are denoted with their Pearson symbols, apart from the common metallic bcc and fcc structures; "dhcp" stands for double hcp. "h–g" stands for a host–guest structure with the "*" after the Pearson symbol denoting the fact that the number of atoms is a non-integer and has been rounded up. Large arrows indicate supposed core ionization (at compression $V/V_o$ equal 0.35 for Li, 0.24 for Na, 0.33 for K, 0.31 for Rb and 0.43 for Cs).

Potassium in its structural sequence under pressure comprises features of the structural sequences of the lighter element Na as well of the heavier elements Rb and Cs. K has however its own special features as for example the absence of a distortive bcc structure $cI$16 observed in the light alkalis Li and Na or related complex structures $oC$52 in Rb and $oC$84 in Cs. These intermediate *post*-fcc structures are missing in K and fcc was observed to transform into a host-guest ($tI$19*) structure or into a hexagonal $hP$4 structure.

Peculiarities in structural behaviour of K under pressure are based on specific features of its electron configuration: K opens in the Periodic table the first long period with 3d transition metals. The empty 3d electron band in K is just above the valence electron 4s level and under pressure there is s – d transition [16]. In heavy alkalis Rb and Cs that have d-levels in their cores electron transitions start at lower pressures than in K. In lighter element Na the 3d level is well above the 3s valence electron level therefore fcc in Na is stable up to a very high pressure of ~100 GPa.

As was considered in [16] the electron configuration of an element essentially changes under strong compression due to broadening and overlapping of electron levels and as result of these changes an electron transfer may occur from core levels into the valence band (i.e. core ionization). Structural transformations of alkali metals from close-packed structures to open-packed and low-coordinated structures indicate convincingly that valence electron configuration is changed. In our recent paper on sodium [17] the question has been raised whether the core ionization takes place for Na in the *oP*8 structure where compression is $V/V_0 \sim 0.25$ and a transition to the divalent state was proposed.

In this paper we investigate possible causes that contribute to the formation of the complex crystal structures in elements under pressure, which have similar crystal structures and atomic volumes for the elements from the left and from the right of the Periodic table. We look at the phase transition sequence of K in details and offer a comparison to other similar structures observed in elements and compounds, as well as suggest the electronic cause for the formation of its crystal structures.

## 2. Theoretical background and method of analysis

The crystal structure of metallic phases is defined by two main energy contributions: electrostatic (Ewald) energy and the electron band structure term. The latter usually favours the formation of superlattices and distorted structures. The energy of valence electrons is decreased due to a formation of Brillouin planes with a wave vector q near the Fermi level $k_F$ and opening of the energy pseudogap on these planes if $q_{hkl} \approx 2k_F$. Within a nearly free-electron model the Fermi sphere radius is defined as $k_F = (3\pi^2 z/V)^{1/3}$, where z is the number of valence electrons per atom and V is the atomic volume. This effect, known as the Hume-Rothery mechanism (or electron concentration rule), was applied to account for the formation and stability of the intermetallic phases in binary simple metal systems like Cu-Zn [18-21], and then extended and widely used to explain the stability of complex phases in various systems, from elemental metals [2,22] to intermetallics, quasicrystals and their approximants [23,24].

The stability of high-pressure phases in potassium are analyzed using a computer program BRIZ [25] that has been recently developed to construct Brillouin zones or extended Brillouin-Jones zones (BZ) and to inscribe a Fermi sphere (FS) with the free-electron radius $k_F$. The resulting BZ polyhedron consists of numerous planes with relatively strong diffraction factor and accommodates well the FS. The volume of BZ's and Fermi spheres can be calculated within this program. The BZ filling by the electron states ($V_{FS}/V_{BZ}$) is estimated by the program, which is important for the understanding of electronic properties and stability of the given phase. For a classical Hume-Rothery phase $Cu_5Zn_8$, the BZ filling by electron states is equal 93%, and is around this number for many other phases stabilized by the Hume-Rothery mechanism.

Diffraction patterns of these phases have a group of strong reflections with their $q_{hkl}$ lying near $2k_F$ and the BZ planes corresponding to these $q_{hkl}$ form a polyhedron that is very close to the FS. The FS would intersect the BZ planes if its radius, $k_F$, is slightly larger then $q_{hkl}/2$, and the program BRIZ can visualize this intersection. One should keep in mind that in reality the FS would usually be deformed due to the BZ-FS interaction and partially involved inside the BZ. The ratio $2k_F/q_{hkl}$, called a "truncation" factor or a "closeness" factor [26], is an important characteristic for a phase stabilized due to a Hume-Rothery mechanism. For Hume-Rothery phases such as $Cu_5Zn_8$, this closeness factor is equal 1.015, and it can reach up to 1.05 in some other phases. This means that the FS radius can be up to approximately 5% larger than the BZ vector $q_{hkl}/2$ for the phase to be stabilized due to a Hume-Rothery mechanism. Thus, with the BRIZ program one can obtain the qualitative picture and some quantitative characteristics on how a structure matches the criteria of the Hume-Rothery mechanism.

The stability of the high-pressure structures of K is analysed here within the nearly-free electron model. This model is particularly valid for alkali metals in its bcc phase as well as in the fcc phase. For the complex phases of K extended Brillouin-Jones zones are constructed with the planes that have a large structure factor, as seen in their diffraction patterns, and lie close to the $2k_F$. The matching criteria for the Hume-Rothery mechanism of the stabilization of these structures (i.e. the number of planes forming the BZ, the closeness factor, and the BZ filling by electron states) are then analysed.

## 3. Results and discussion

Characteristics of crystal structures for potassium at ambient and high pressure are given in Table 2 as published in the literature (the incommensurate host-guest structure K-*tI*19* is discussed later in *3.4.*). Our considerations of these structures within the BRIZ program [25] are given in Table 2 representing the $k_F$ values for the corresponding numbers of valence electrons, characteristics of Brillouin zone (BZ) planes which are close to the Fermi sphere (FS), degree of overlapping BZ planes with the ideal FS and of calculated filling of Brillouin zones by electron states $V_{FS}/V_{BZ}$. Close-packed structures bcc and fcc, common for metals, transform in potassium to open-packed, low symmetry structures with coordination number decrease from 8+6 for bcc and 12 for fcc to 9-10 for *tI*19*, 6-8 for *hP*4, ~8 for *oP*8 and *tI*4 with further increase to 10-11 for *oC*16. At first breakpoint in coordination number near 20 GPa where V/V$_o$ equal 0.33 for K valence band overlap with the core electrons is proposed implying that K is no more a monovalent metal (estimation is shown in the Table 2).

**Table 2.** Structure parameters of K phases as given in the literature. The Fermi sphere radius k$_F$, the total number of Brillouin zone planes, the ratio of 2k$_F$ to Brillouin zone vectors ($2k_F/q_{hkl}$) and the degree of filling of Brillouin zones by electron states V$_{FS}$/V$_{BZ}$ are calculated by the program BRIZ [25].

| Phase | K-bcc | K-fcc | K-*hP*4 | K-*oP*8 | K-*tI*4 | K-*oC*16 |
|---|---|---|---|---|---|---|
| **Structural data** | | | | | | |
| Pearson symbol | *cI*2 | *cF*4 | *hP*4 | *oP*8 | *tI*4 | *oC*16 |
| Space group | $Im\bar{3}m$ | $Fm\bar{3}m$ | $P6_3/mmc$ | $Pnma$ | $I4_1/amd$ | $Cmca$ |
| P,T conditions | Ambient conditions | 11.7GPa T = 300K | 25 GPa T=300K | 58 GPa T=300K | 112 GPa | 112 GPa |
| Lattice parameters (Å) | $a$ = 5.321 | $a$ = 5.130 | $a$ = 4.218 $c$ = 5.737 | $a$ = 5.602 $b$ = 3.311 $c$ = 6.380 | $a$ = 2.322 $c$ = 8.669 | $a$ = 8.03 $b$ = 4.753 $c$ = 4.716 |
| Atomic volume (Å$^3$) V/V$_0$ | 75.327 1 | 33.751 0.448 | 22.10 0.293 | 14.79 0.196 | 11.68 0.155 | 11.25 0.149 |
| Atomic positions | 2a (0,0,0) | 4a (0,0,0) | 2a (0,0,0) 2c (⅓,⅔,¼) | 4c ($x_1$,¼,$z_1$) 4c ($x_2$,¼,$z_2$) $x_1$ = 0.034 $z_1$ = 0.156 $x_2$ = 0.144 $z_2$ = 0.567 | 4a (0,0,0) | 8d (x, 0, 0) x = 0.2161 8f (0, y, z) y = 0.173 z = 0.327 |
| Reference | [27] | [28] | [9] | [8] | [8] | [8] |
| **FS - BZ data from the BRIZ program** | | | | | | |
| z (number of valence electrons per atom) | 1 | 1 | 2.5 | 2 | 2.5 | 4 |
| k$_F$ (Å$^{-1}$) | 0.733 | 0.957 | 1.496 | 1.588 | 1.850 | 2.192 |
| Total number of BZ planes | 12 | 14 | 18 | 34 | 16 | 32 |
| HKL and 2k$_F$/q$_{hkl}$ | *110*: 0.877 | *111*: 0.902 *200*: 0.781 | *102*: 1.074 *110*: 1.004 | *210*: 1.081 *112*: 1.074 *211*: 1.023 *103*: 1.005 *301*: 0.906 *020*: 0.837 | *103*:1.066 *112*: 0.904 | *131*: 1.0301 *113*: 1.0238 *421*: 1.0176 *511*: 1.0103 |
| Filling of BZ with electron states V$_{FS}$/V$_{BZ}$ (%) | 50 | 50 | 93 | 93.2 | 79.2 | 93 |

*3.1. Compact structures (bcc, fcc)*

First, we consider the compact structures experimentally observed in potassium (as well as in other alkali metals), that are closely packed, have high coordination number and high Madelung constant, i.e. bcc and fcc. The bcc structure has the highest Madelung constant α = 1.791858, followed by fcc with α = 1.791747 and hcp with α = 1.791676 [29] though the packing of atoms is the highest for fcc and hcp (with the ideal c/a) and equals 0.74 compared to 0.68 for bcc. The electrostatic Madelung energy can be defined by Ewald-Fuchs method in terms of lattice sums in real and reciprocal space [29]. A qualitative argument for the highest value of the Madelung constant for the bcc structure can be evident from the Brillouin zone configuration consisting of 12 equidistant planes that is more favourable than the 8+6 planes for the fcc structure (Figure 3). Thus the stability of the bcc structure of K (and other alkalis) at ambient pressure can be attributed to the electrostatic energy.

The bcc and fcc structures are closely-packed metallic phases and the bcc – fcc transformation occurs without remarkable volume change. Thus no discontinuous volume decrease was observed at the K-I → K-II transition at 11 GPa [30] and for Na the relative volume difference $(V_{fcc}-V_{bcc})/V_{bcc}$ is about 0.1% [31]. For Li at the bcc-fcc transition near 7.5 GPa (298 K) the difference is also small - about 1.6% [32].

Transition of alkali metals under pressure from bcc to fcc can be regarded as an increase in the other contribution to crystal energy – the band structure energy – trough Brillouin plane -Fermi surface interaction and appearance of FS necks toward the BZ planes (see for example [33]). Adopting the fcc structure alkali metals become more similar to the group-IB elements (Cu, Ag and Au) known as noble metals. The ground state of these elements is fcc which is stable to very high pressures. In Au it is stable to as high as ~250 GPa where at compression of $V/V_0 \approx 0.65$ it is shown to transform to hcp [34].

Thus, the group-IA elements (alkalis) and group-IB (noble metals) having formally one valence electron of s-type display considerably different behaviour under compression. This difference is defined by different electron configuration of these group elements: for noble metals the upper d-band is filled (3d in Cu, 4d in Ag and 5d in Au) whereas the same d-bands are empty for alkalis (K, Rb and Cs) (see textbook Ashcroft & Mermin [35], Chapter 15)]. Electron configurations in noble metals with the filled d-band are attributed to their low compressibility in contrast to the very high compressibility of alkali metals. Under pressure the empty d-bands in the alkali metals go down and overlap with the outer s-electron band and on further compression also with the core electron bands which results in the open-packed and complex crystal structures.

This remarkably different response to compression for these two groups of monovalent metals implies a significantly different change in electron band structure energy for alkali metals compared to noble metals. The earlier suggested transition of s electron to the empty d band that lowers under pressure (the so called s – d transfer) on its own is not enough to explain the observed variety of structures in alkalis under pressure, therefore a more essential reconstruction of the electron band structure in compressed alkalis should be considered. It was proposed that Na in the high-pressure *oP*8 structure should contain 2 valence electrons which assumes an overlap of outermost core electrons with the valence band [17].

Analysing the *post* - fcc structures for all alkali metals, we can see for the lighter alkali metal Li the formation of a distorted fcc structure – the rhombohedral *hR*1 phase [36], which is similar to α-Hg. This distortion of fcc appears due to the increase of the electron band structure energy over the electrostatic term on compression. Another kind of fcc distortion – from a cubic to a tetragonal cell – was found in Pb-In alloys under pressure driven by same mechanism [37]. There are no electronic transitions connected to these transitions of the fcc phase neither to the formation of the *cI*16 phase in Li and Na at higher pressures that can be considered as supercell of bcc. The origin of stability of the *cI*16 structure is a reduction in the electronic energy due to formation of new Brillouin zone planes close to the Fermi sphere [2]. Related to *cI*16, more complex crystal structures were found in heavy alkalis Rb-*oC*52 and Cs-*oC*84 [1] where additional complexity is commonly attributed to the electron s – d transfer and was explained within the BZ-FS interactions [2]. Potassium is the only alkali element where the fcc structure transformed directly to open-packed and complex structures, either *tI*19* or *hP*4, discussed below.

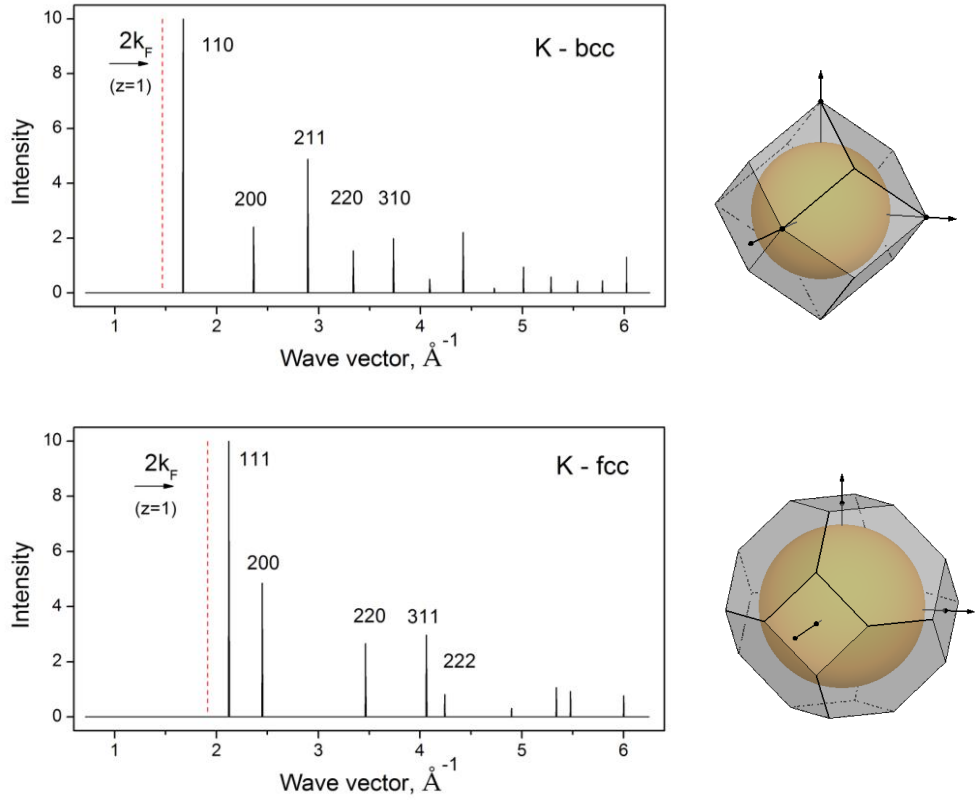

Figure 3. Calculated diffraction patterns of potassium (left) and corresponding Brillouin zones with the inscribed FS (right). In the top panel – for K-bcc at the ambient pressure and in the bottom panel – for K-fcc at 12 GPa with structural parameters given in Table 2. The position of $2k_F$ for given z and the *hkl* indices of the planes are indicated on the diffraction patterns.

*3.2. K-hP4 of NiAs-type and K-oP8, distortion of NiAs*

Further compression of K above 20 GPa results in transitions from typically metallic close-packed fcc structure to open-packed and low-coordinated structures either host-guest *tI*19[*] or *hP*4. The K-*tI*19[*] is discussed below (in *3.4.*) and here we focus on K-*hP*4, observed between 25 and 35 GPa in two different samples [9]. The NiAs-type *hP*4 structure has the coordination number (CN) equal to 6 and 6+2 which is a great decrease in density of packing compared to CN=12 for fcc. Interatomic distances ($d_{min}$) for K-*hP*4 at 25 GPa are 2.826 and 2.868Å (CN=6, 6+2) that are close to the double ionic radius $2r_i$ = 2.66Å ($r_i$=1.33Å given by Kittel [38]). It is necessary to compare $d_{min}$ for K-*hP*4 with data of ionic radii estimated by Shannon [39] corrected for coordination number (given in brackets) stated as 1.38(6), 1.51(8), 1.59(10) which result in $2r_i$ equal to 2.76, 3.05, 3.18Å. This comparison shows that interatomic distances for K-*hP*4 are similar to or *smaller* than the double ionic radius $2r_i$ = 3.05Å (for CN=8) which leads to conclusion that at compression above 20 GPa ions of K should overleap and the upper core electrons should participate in the valence bonding. In accordance with this suggestion the number of valence electrons is proposed here for all *post*-fcc structures to be more than 1 and equal to 2 (for *oP*8) or 2.5 (for *hP*4).

Decrease in coordination number is accompanied by a drastic decrease in atomic volume as found experimentally. The relative volume change ($V/V_{trans}$) at the K-fcc→K-*tI*19* transition is about 10% as estimated from Figure 2. Atomic volume for K-*hP*4 is nearly equal to that of K-*tI*19* (and this may be the reason for observation of these two phases at same pressures in different samples). The drop in atomic volume with decrease of packing density is considered to be a sign of essential decrease in atomic radius as the result of core ionization.

The next phase, K-*oP*8, exists at pressures 54 – 90 GPa with CN=8 and minimal interatomic distances 2.46 - 2.51 Å which is essentially smaller than $2r_i$ (= 3.05Å for CN=8). Structural data for K-*hP*4 and K-*oP*8 are given in Table 2. Apart from K, the *oP*8 structure was found in only one other element – Na – at pressures of 117-125 GPa as indicated in Table 1. Structural sequences for these elements, Na and K, are different and Na-*hP*4 was observed at ~2 Mbar after the host-guest structure. This difference is due to different electronic levels in these atoms: Na is element of the 3$^{rd}$ row in the Periodic Table with the empty 3d band well above the occupied 3s band and K is in 4$^{th}$ row with the 3d band close above the occupied 4s band. The overlap 4s – 3d bands occurs in K under compression to ~20GPa, whereas fcc-Na is stable to much higher pressure of ~100 GPa.

*3.3.1. hP4 and oP8 – the Hume-Rothery phases.* Now using the FS-BZ concept, we analyse the stability of the high-pressure phases of potassium with the *hP*4 and *oP*8 structures. Diffraction patterns of these phases and constructions of Brillouin zones and inscribed Fermi spheres are shown in Figure 4. Stability of these structures in potassium can be qualitatively understood within the Hume-Rothery mechanism assuming the number of valence electrons equal to 2.5 and 2 for *hP*4 and *oP*8, respectively. This values coincide with the number of valence electrons per atom in the known binary simple metal compounds AuSn-*hP*4 (z=2.5) and AuGa-*oP*8 (z=2) [40].

Brillouin zones considered for these structures consist of many planes in close contact with the Fermi sphere that is overlapping with the planes within this idealised model, however would be involved inside the zone in the real case of a non-spherical and distorted Fermi surface. The BZ filling by electron states is high – up to ~80% for *hP*4 and up to 92% for *oP*8 which should be compared with 93% for the classical Hume-Rothery phase $Cu_5Zn_8$- *cP*52.

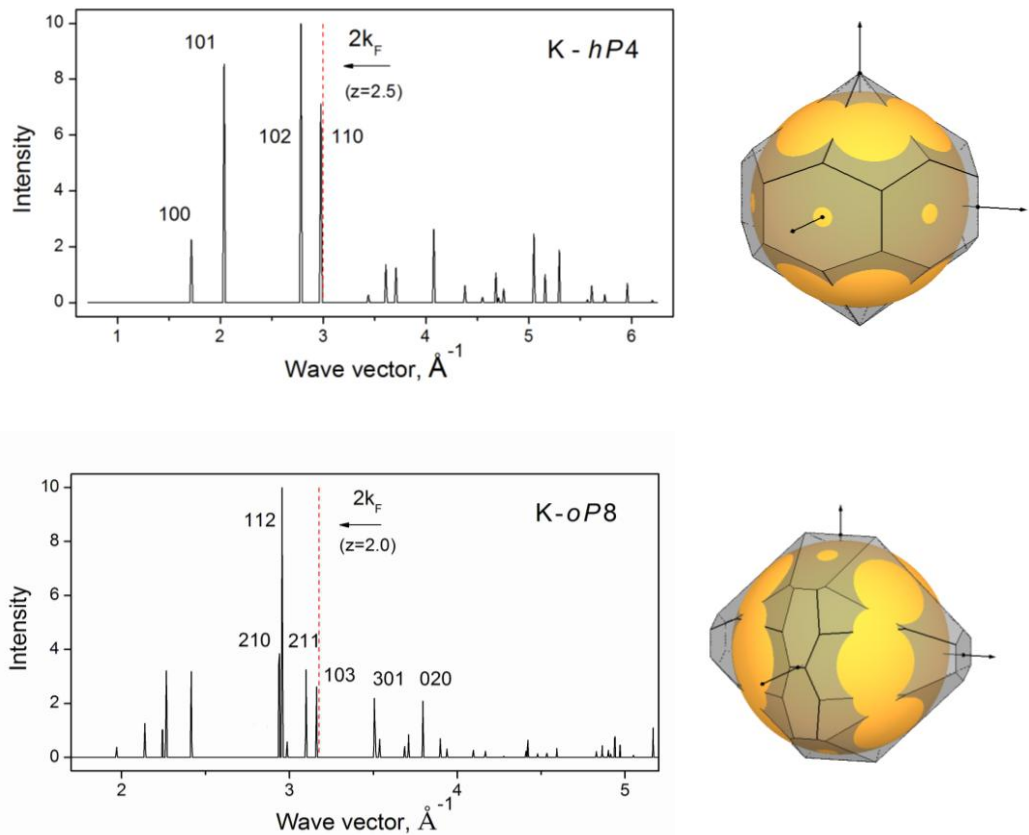

Figure 4. Calculated diffraction patterns of potassium (left) and corresponding Brillouin zones with the inscribed FS (right). In the top panel – for K- *hP*4 and in the bottom panel – for K- *oP*8 with structural parameters given in Table 2. The position of 2$k_F$ for given z and the *hkl* indices of the main planes are indicated on the diffraction patterns.

*3.3.2. Structural relation of hP4 and oP8.* The *oP*8 MnP-type structure is an orthorhombic distortion of the hexagonal *hP*4 NiAs-type as was discussed in details in our paper on sodium [17]. Relation of the orthorhombic (o) and hexagonal (h) lattices is as follows: $a_o ≈ c_h$, $b_o ≈ a_h$, and $c_o ≈ a_h\sqrt{3}$. In the *oP*8 structure, atoms are slightly displaced from the ideal positions of the *hP*4 structure. Atomic positions in the *Pnma* space group found for K and Na are very similar to that of known 20 binary phases with *oP*8 MnP-type structure [41]. The structural transformation from *hP*4 NiAs-type to *oP*8 MnP-type is accompanied by an increase in coordination number from 6 and 8 in *hP*4 to 10 and 8 in *oP*8 for two different atomic sites.

The formation of the *oP*8 structure from *hP*4 by distortion of lattice parameters with some atomic displacements leads to the appearance of numerous additional diffraction peaks as seen in Figure 4. Therefore the Brillouin-Jones zone for *oP*8 has a number of additional planes compared to the BZ of *hP*4 and a reduced volume of the BZ, which leads to an increase in the BZ-FS interaction and the lowering of the electronic energy.

Phase transitions between *hP*4 and *oP*8 phases are known as a function of temperature and composition [42,43]. The transformation NiAs-type (*hP*4) to MnP-type (*oP*8) was found at high pressure in the FeS compound [44]. There are more examples demonstrating the phase transformation under pressure from NiAs-*hP*4 to MnP-*oP*8 as discussed in our paper on Na-*oP*8 [17]. If one considers compounds with simple metals as constituent elements having sp type of valence electrons, the *hP*4 and *oP*8 phases exist within the effective number of valence electrons (z) between 2 and 2.5 as for example AuGa-*oP*8 (z=2) and AuSn-*hP*4 (z=2.5) [40].

The relation of the *oP*8 structure to the *hP*4 NiAs type can be extended to a wider structural group of hexagonal phases with the common *Structurbericht* type B8 that combines $Ni_2In$ and NiAs structure types; the former is a filled up type of the latter. The number of the known B8-type phases has been growing steadily to a today total of more than 400 compounds giving "examples of diverse structural behaviour" (superstructures, ordering of vacancies etc) [45]. Stability of binary phases related to B8-type structure was considered within the Hume-Rothery mechanism [46]. The subjects of consideration are phases in the binary systems of a noble metal and an element from the main groups of the Periodic Table. This allowed us to define the phases using a certain value of electron concentration and to apply Hume-Rothery arguments to the stability of these complex phases.

*3.3.3. Electron configuration in K-hP4 and K-oP8.* The *oP*8 structure found in Na and K has no structural analogues among elements. A search for some binary analogues yielded a comparison of this structure to an anti-cotunnite crystal structure ($PbCl_2$ is a prototype of cotunnite and a representative example of anti-cotunnite is $Co_2Si$). It was suggested for alkali elements that at strong compression electrons are forced away from the near core regions resulting in an increased electron density in the *interstitial* regions [47-49]. This interpretation is within the concept of relating high-pressure forms of alkali metals to certain binary compounds, so called "electride" compounds, with alkali metal atoms occupying the cation positions and electron density maxima located in the empty anion positions. As shown above, the *hP*4 and *oP*8 structures in Na and K are crystallographically related to the family of the *Structurbericht* B8 type containing the $Ni_2In$-type (*hP*6) and NiAs-type (*hP*4) structures and through an orthorhombic distortion to the $Co_2Si$-type (*oP*12) and MnP-type (*oP*8) structures. The consideration of the *hP*4 and *oP*8 phases as an 'electride' relates them to the $Ni_2In$(*hP*6) and $Co_2Si$ (*oP*12) structures with atoms Na or K occupying the cation atomic positions and the interstitially localized electrons occupying empty anion positions. Potassium under pressure is treated as "a pseudobinary ionic compound" [9].

In present work as in our previous paper on Na [17], an *alternative approach* is suggested for the interpretation of the *hP*4 and *oP*8, in which K atoms occupy the atomic positions of AuSn of the NiAs-type (*hP*4) and AuGa of the MnP-type (*oP*8).These phases are considered as Hume-Rothery phases stabilized at the certain number of valence electrons per atom (2.5 in the *hP*4 and 2 in the *oP*8). This interpretation has led us to the assumption that alkali metals Na and K in these structures are no longer monovalent metals becoming a divalent metal or even of higher valency due to the core overlap with the conduction band.

*3.3. K-tI4 and K-oC16 structures*

The body-centered tetragonal structure, *tI*4, space group $I4_1/amd$, is unique to the alkali metals and was found previously for Cs above 4.3 GPa [50] and for Rb above 20 GPa [28]. The *oC*16 structure, C-face centred orthorhombic, space group *Cmca*, was observed in alkali elements Cs and Rb [51,52] as well as in the group IV elements Si and Ge [53,13]. Recently this structure was determined for the group V element - bismuth at 3.2GPa, 465K [54] and have been reported for Bi-based alloys in the earlier papers [55,56].

Experimental studies of K to 112 GPa [8] give the volume changes (V/$V_{trans}$) for the *oP*8→*tI*4 and *tI*4→*oC*16 transitions at 90 and 96 GPa as ~1% and 3.7%, respectively. The *oP*8→*tI*4 transition is not known in any other alkali metals. The volume changes at the $tI4 \rightarrow oC16$ transition in Rb and Cs are 3.3% and 9.3%, respectively [52,51]. Coordination number along the transitions $oP8 \rightarrow tI4 \rightarrow oC16$ changes as 8,10 → 4+4 → 10,11. Thus at the $oP8 \rightarrow tI4$ transition there is a decrease in coordination number and very small decrease in atomic volume. The transition $tI4 \rightarrow oC16$ is accompanied with increase in coordination number and significant decrease in atomic volume that implies possible considerable changes in the valence electron bonding.

Commonly accepted explanation of the stability of $tI4$ and $oC16$ phases in alkali metals consists in the s – d transfer and the formation of electride-like compounds with high electron density in the anion positions. For the same $oC16$-$Cmca$ phases in Si and Cs theoretical calculations found quite different behaviour – a nearly-free electron behaviour in Si in contrast to "significant electron densities" located in the interstitial positions in Cs [57].

We propose an alternative description for the $tI4$ and $oC16$ structures of K that follow the $oP8$ under pressure and suggest that these transitions are accompanied by further decrease of ionic radius and increase in the number of valence electrons. In Figure 5 are shown diffraction patterns for potassium structures $tI4$ and $oC16$ and constructed Brillouin zones with inscribed Fermi spheres for the number of valence electrons 2.5 and 4, respectively.

As can be seen from Figure 2, phases K-$oC16$ and Ge-$oC16$ exist at pressure ~100 GPa having nearly same atomic volumes and lattice parameters [8,13]. This observation formed a ground for the assumption that these phases have the same number of valence electrons per atom equal to 4. The Brillouin zone constructed for this structure has many planes (32 as indicated in Table 2) in close contact with the Fermi sphere and the BZ filling by electron state is 93% as for the γ-brass Hume-Rothery phase (see discussion in [2]).

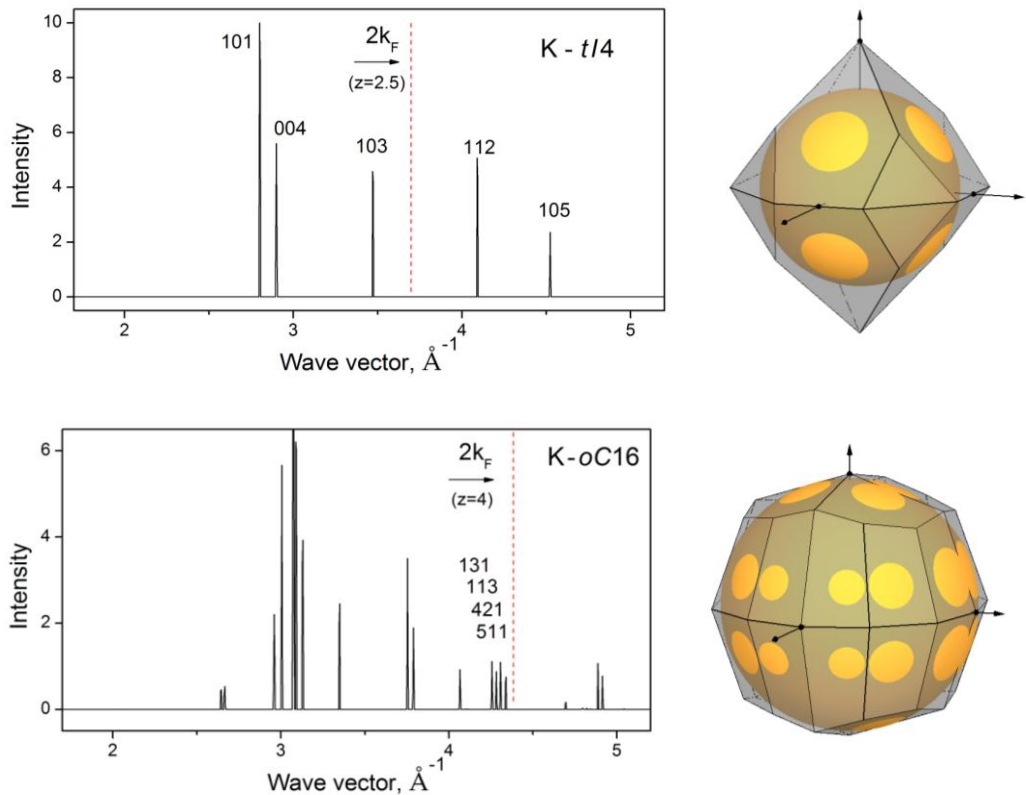

Figure 5. Calculated diffraction patterns of potassium (left) and corresponding Brillouin zones with the inscribed FS (right). In the top panel – for K- $tI4$ and in the bottom panel – for K- $oC16$ with structural parameters given in Table 2. The position of $2k_F$ for given z and the *hkl* indices of the main planes are indicated on the diffraction patterns.

*3.4. Incommensurate host-guest structure K-tI19\**

Potassium transforms at ~20 GPa from fcc to the host-guest *tI*19* structure [30] with a decrease in coordination number from CN=12 to CN=9,10 and with a significant drop in atomic volume (about 10%) as considered in *3.2*. The minimal interatomic distance $d_{min}$ for K - *tI*19* is 2.832 Å which is *smaller* than the double ionic radii $2r_i$ = 3.05Å (for CN=8) and 3.18Å (for CN=10) estimated by Shannon [39] This implies for potassium an increase in the number of valence electrons per atom in the first *post*-fcc phase – K-*tI*19*.

Similar incommensurate host-guest structure *tI*19* was found in Rb and Na with the difference in the guest subcell (K: C-face-centered, Rb: body-centered, Na: monoclinic) [58,59]. The other type of host-guest structure assigned as *tI*11* is formed under pressure in Ba, Sr, Sc, Bi, Sb and As (see reviews [1,3]) and recently found in Ca [60]. We found structural relationship between atomic arrangement in these two types of host-guest structures and a binary simple metal phase $Au_3Cd_5$-*tI*32 [61]. Phases $Au_3Cd_5$ and $Au_5Cd_8$ exist on the phase diagram Au-Cd in the same composition range (~62 at% Cd) differing only by temperature: $Au_3Cd_5$ transforms to $Au_5Cd_8$ above 370°C. The crystal structure of $Au_5Cd_8$ is of $Cu_5Zn_8$-*cI*52-type well known as a Hume-Rothery phase stabilized by a Fermi sphere – Brillouin zone interaction (see [19,24]). Therefore stability of $Au_3Cd_5$-*tI*32 can be considered within the Hume-Rothery mechanism.

In the $Au_3Cd_5$-*tI*32, the arrangement of Au and Cd atoms in the space group *I4/mcm* is shown in Figure 6 for the layer at z=0. Cd atoms in position 16k form octagon (eight-sided polygon) - square nets same as K-host atoms in *tI*19*. Au atoms in position 8h form square - triangle nets like Bi-host atoms in *tI*11*. Atoms Cd in 4b and Au in 4a relate to guest atoms in chains in K and Bi host-guest structures with an increased interatomic distances where a guest subcell is formed with an incommensurate $c_{host}/c_{guest}$ ratio and the number of atoms in the structure is reduced to non-integer values

Axial ratio for $Au_3Cd_5$-*tI*32 is equal to c/a =0.499 and is comparable to $(c/a)_{host}$ for *tI*19* equal to 0.485 for K at 20GPa, 0.491 for Na at 130GPa and 0.501 for Rb at 16.8GPa. For *tI*11* in Bi $(c/a)_{host}$ = 0.489 also close to ~0.5, whereas in Ba and Sr $(c/a)_{host}$ is considerably higher (0.562 and 0.569, respectively). For our consideration we choose the structures $Au_3Cd_5$-*tI*32, K-*tI*19* and Bi-*tI*11*. Characteristics of the considered structures are given in Table 3.

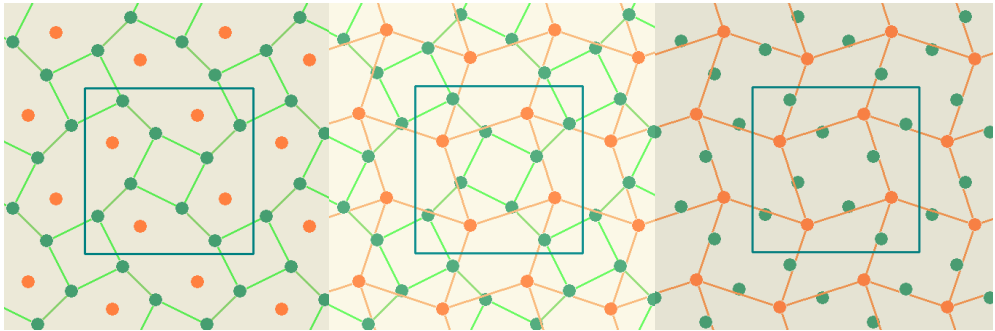

Figure 6. Atomic configuration for $Au_3Cd_5$-*tI*32, *I4/mcm* in the layer at z=0 (middle).
Cd atoms (green/dark gray) in position 16k form octagon - square nets like K-host in *tI*19 (left).
Au atoms (red/gray) in position 8h form square - triangle nets like Bi-host in *tI*11* (right)

**Table 3.** Structure parameters of host-guest phases K-*tI*19*, Bi-*tI*11* and Au$_3$Cd$_5$-*tI*32 as given in the literature. The Fermi sphere radius k$_F$, the total number of Brillouin zone planes, the ratio of 2k$_F$ to Brillouin zone vectors (2$k_F$/$q_{hkl}$) and the degree of filling of Brillouin zones by electron states V$_{FS}$/V$_{BZ}$ are calculated by the program BRIZ [25].

| Phase | K-(h-g)*tI*19* | Bi-(h-g)*tI*11* | Au$_3$Cd$_5$ |
|---|---|---|---|
| **Structural data** | | | |
| Pearson symbol | (h-g)*tI*19* | (h-g)*tI*11* | *tI*32 |
| Space group | *I4/mcm*(00γ)000*s* | *I'4/mcm*(00γ)0000 | *I4/mcm* |
| P,T conditions | 22.1GPa T=300K | 6.8 GPa T=300K | 0 GPa T=300K |
| Lattice parameters (Å) | $a_H$ = 9.767 $c_H$ = 4.732 $c_G$ = 2.952 γ = 1.603 | $a_H$ = 8.518 $c_H$ = 4.164 $c_G$ = 3.180 γ = 1.309 | a = 10.728 c = 5.352 c/a = 0.499 |
| Atomic volume (Å$^3$) V/V$_0$ | 23.50 0.312 | 28.326 0.80 | 19.25 |
| Atomic positions | Host atoms ($x_H$, $y_H$, ½) Guest atoms (½,0,0) $x_H$ = 0.79 $y_H$ = 0.085 | Host atoms ($x_H$, $y_H$, 0) Guest atoms (0,0,0) $x_H$ = 0.153 $y_H$ = 0.553 | Au 8*h* (0.167, 0.667, 0) Au 4*a* (0, 0, ¼) Cd 16*k* (0.078, 0.225, 0) Cd 4*b* (0, ½, ¼) |
| Reference | [30,1] | [62] | [61] |
| **FS - BZ data from the BRIZ program** | | | |
| z (number of valence electrons per atom) | 2.6 | 4.75 | 1.625 |
| $k_F$ (Å$^{-1}$) | 1.485 | 1.704 | 1.357 |
| Total number of BZ planes | 48 | 48 | 32 |
| HKL and 2$k_F$/$q_{hkl}$ | 420: 1.0323 3101: 1.0097 202: 1.0065 411: 1.0013 | 420: 1.0331 3201: 1.022 202: 1.0145 411: 1.004 | 420: 1.0362 202: 1.0283 411: 1.0108 |
| Filling of BZ with electron states V$_{FS}$/V$_{BZ}$ (%) | 95 | 95 | 95 |

Interestingly, that for the host-guest structures in Ba and Rb "a remarkable duality" was found when the representation of structure is considered in terms of atoms and localized electrons ("blobs") [63]. The appearance of the interstitial electron blobs has been described as the formation of an `electride' in which the interstitial electrons are the anions. The Ba-IV and Rb-IV structures are therefore related "by reversing the roles of the ions and electron blobs" [63].

In our consideration we propose an overleap of upper core electrons with the valence band implying an increase in the number of valence electrons and involving the Hume-Rothery mechanism for understanding of complex phase formation. Figure 7 (middle) shows diffraction pattern and the constructed Brillouin zone with the planes close to the Fermi sphere counting the number of valence electrons per atom z = 1.625. The planes 420, 202 and 411 form a polyhedron with 32 planes accommodating well the Fermi sphere with the filling by electron states equal to 95% which meets the Hume-Rothery mechanism criteria.

To work out the number of valence electrons for K and Bi in their host-guest phases we use the following idea that the number of valence electrons in a Brillouin zone for the same Bravais lattice is held nearly constant: $zN \approx$ const, were N is the number of atoms in the cell, e.g. if the number of valence electrons increases, the number of atoms in the cell would decrease to keep the zN nearly constant. It was suggested by Jones (see ref. [19] p.206) that "the large zone in k-space is defined by the Bravais lattice on which the structure is based and the Miller indices of its faces. Hence its volume and the number of states it contains, is independent of the number of atoms in the unit cell." To a first approximation, the Brillouin-Jones zone can be used to estimate the likely electron count for a participating element. For $Au_3Cd_5$ where z=1.625 and N=32 we have zN = 52. Taking the same Bravais lattice and the same Brillouin zone we obtain for K-*tI*19* with N=19.2  z = (52/19.2) = 2.7 and for Bi-*tI*11* with N=10.6  z = (52/10.6 ) = 4.9.

The valence electron number per atom for K and Bi given in Table 3 is reduced to comply with degree of BZ filling ~ 95% and accepted to be 2.6 for K-*tI*19* and 4.75 for Bi-*tI*11*. Non-integer electron-per-atom counts appear due to overlap of *s* (or *sp*) and *d* levels and within the FS-BZ model only *s* (or *sp*) electrons are taken into account as structure controlling electrons.

For Bi-*tI*11* host atoms (8h) are assumed to have 5 valence electrons per atom and guest atoms (there are 2.6 guest atoms in the host unit cell) have 4 electrons per atom giving the average electron count 4.75 per atom. The decrease in z from 5 to 4 for guest Bi atoms is assumed in the above consideration and is possible because a lowering of the upper empty *d*-level may occur with increasing temperature or pressure which would lead to the level hybritisation. Following the same arguments, one can rationalize the appearance in Bi of the *oC*16 structure with the stability range around z=4 electrons per atom which was found to appear in Bi at same pressures as *tI*11* but at higher temperature. Similar, for K-(h-g) *tI*19* one can expect different valency for the host and guest atoms with more participation of *d*-electrons for guest atoms that provides strong bonding in chains of guest atoms with short distances and gives rise to effect of chain melting [64] and different guest-atom-chains arrangements [10].

Diffraction patterns for host-guest structures K-*tI*19* and Bi-*tI*11* are shown in Figure 7 with indication of $2k_F$ position for corresponding z. The BZ polyhedrons are constructed for commensurate approximants assuming γ equal 8/5=1.6 and 4/3=1.333, respectively. BZ's for h-g structures contain additional planes from guest reflections compared to BZ for $Au_3Cd_5$ that reduce slightly the volume of BZ leading to nearly full BZ. Therefore electron counts were defined from the 95% filling of the zone resulting in z=2.6 for K and 4.75 for Bi. The total number of BZ planes in the contact with the FS increases to 48 (see Table 3) that should result on the substantial changes of physical properties at the transition to these host-guest structures.

Both *tI*11* and *tI*19* incommensurate host-guest structures form additional planes at the Fermi sphere compared to a commensurate basic structure leading to the gain in the electron band structure energy even if this is not favourable for the electrostatic term that prefers more symmetrical arrangement. This shift of balance in the structural energy may account for incommensurate structures of other types. For example, found in phosphorus above 100 GPa, P-IV, an orthorhombic structure with an incommensurately modulation wave vector along the c-axis was characterized by appearance of satellite reflections and hence by a formation of additional Brillouin zone planes near the FS [65].

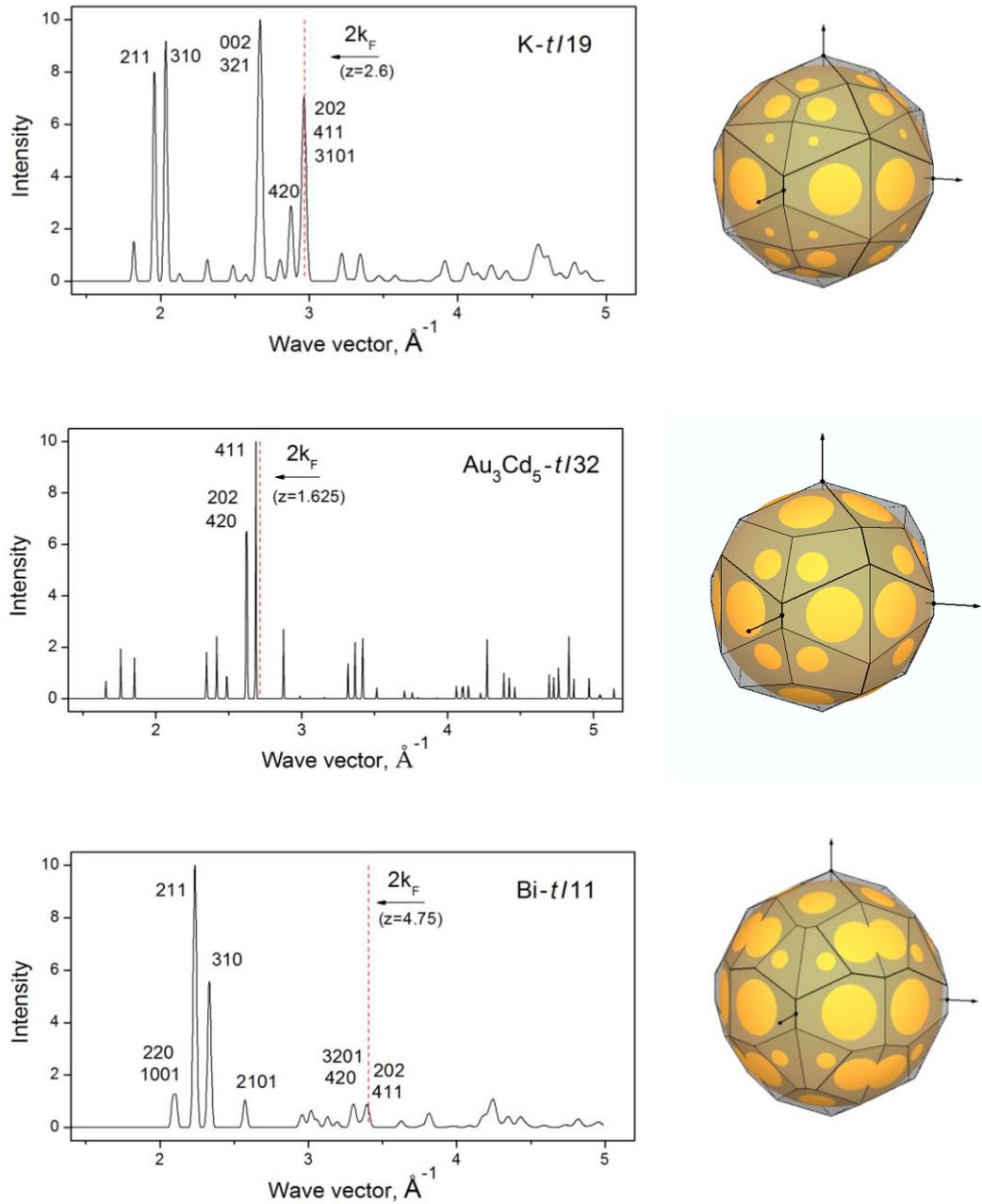

Figure 7. Calculated diffraction patterns (left) and corresponding Brillouin zones with the inscribed FS (right). In the top panel – for K- $tI$19*, in the middle panel – for $Au_3Cd_5$-$tI$32 and in the bottom panel – for Bi- $tI$11* with structural parameters given in Table 3. The position of $2k_F$ for given z and the *hkl* indices of the principal planes are indicated on the diffraction patterns.

*3.5 Structure – properties correlations in post-fcc K: core ionization*

Experimental measurements of electrical resistance and optical reflectivity of potassium under pressure have shown considerable changes of these properties upon phase transitions [66,67]. Resistance of K exhibits a rise by a factor of 50 in the pressure range up to 60 GPa and this behaviour was suggested to occur "when core ionization takes place" [68]. Decrease in optical reflectivity found in [67] gives "a clear evidence for an optical gap over extended regions of the Brillouin zone and for a large density of occupied states close to $E_F$." Both bcc and fcc structures of alkali metals are characterized as good metals and a significant rise of resistively and loss of reflectivity for K at the transition fcc – *tI*19* can be understood due to essential difference in the BZ-FS configuration and the degree of BZ filling by electron states estimated as 0.50 to 0.95, respectively.

The low reflectivity of Rb-IV (host-guest) was found in [69] which is in alignment with the 95% BZ filling proposed for the host-guest K-III structure. The transition in sodium to host-guest phase Na-*tI*19* at 125 GPa is also accompanied by a marked decrease in optical reflectivity [58]. Authors [69] assigned the changes in reflectivity to "structural transitions that may be accompanied by significant changes of the complicated Fermi-surface topology".

Measurements of electrical resistance have shown for the transition *tI*19* – *tI*4 in Rb a substantial decrease and an increase for *tI*4 – *oC*16 transition in Cs [70]. One can expect for potassium from structural analogy with the other alkali metals along transitions *tI*19* – *tI*4 – *oC*16 the same changes in resistivity – after observed increase in *tI*19* one can expect to see a drop in resistivity at the transition to *tI*4 and again increase in resistivity at the transition to *oC*16. This behaviour should traced the Brillouin zone – Fermi sphere configuration and degree of BZ filling estimated as 0.95 – 0.79 – 0.93, respectively (see Tables 2 and 3). Correlations of physical properties and crystal structures of alkali elements can be understood by assuming core ionization under compression.

There is a similarity with the effect of *chemical* pressure that can arise in alloys of alkalis with transition and post-transition metals. The difference in atomic volume for these components is significant to provide a compression of alkali metals by formation of a compound. An interesting example is the $Ag_5Li_8$-*cI*52 phase with the structure of γ - brass usually attributed to electron concentration 21/13. Hume-Rothery, considering the $Ag_5Li_8$ phase in his textbook (see ref. [20] p.306] suggested that its stability will satisfy the electron concentration 21/13 as for other γ - brass phases, if lithium were divalent.

Another compound $K_2Ag$ with the structure related to the Hume-Rothery type $CuZn_2$-*hP*3 was found in the K-Ag alloys under pressure ~6 GPa [71], "where the distances between the K atoms within the graphite-like layers at 6.1 GPa are 3.13 Å, comparable to those in elemental potassium at 44 GPa". This structure in $CuZn_2$ stabilizes by the Hume-Rothery mechanism at z ≈ 1.7 electrons per atom which can be applied to $K_2Ag$ if K were divalent at such compression. Considerations of $CuZn_2$-*hP*3 and related phases within the Brillouin zone - Fermi sphere interaction were given recently [46] where the family of structures related to $CuZn_2$-*hP*3 is stabilized by the Hume-Rothery electron concentration rule. In accordance with this condition the stability of other known compounds such as $Li_2Pd$ or $Li_2Pt$ [40] should be considered with the valency of alkali metals higher than 1.

**4. Summary and conclusions**

Structural transformations of potassium are considered in comparison with other alkali elements and elements from groups IVB and VB, in particular with Ge and Bi. Complex low-symmetry crystal structures are discussed that were found in K above 20 GPa just after fcc with significant reduction in atomic volume. Commonly accepted explanation of these transformations within s – d electron transfer is not sufficient to describe the structural diversity observed in alkalis.

It is useful to look for relations between the structures of compressed alkali metals and some binary phases and to consider the main factors responsible for the formation of complex phases in simple metals and alloys, in the first place the Hume-Rothery mechanism. Structures *hP*4 and *oP*8 found in K have a similarity with binary phases AuSn and AuGa which are characterised by the electron concentration 2.5 and 2, respectively. We suggest that in potassium at densities higher than $V/V_o ≈ 0.3$ there is not only s – d electron transfer but also core electrons start to participate in the valence band and K becomes a polyvalent metal with the number of valence electrons counting to 2 in *oP*8 and even to 4 in *oC*16. Interatomic distances for K in these structures are shown to be smaller than the double ionic radius that supports the assumption of core ionization.

Alkali metals on compression adopt the open-packed and complex crystal structures following in opposite direction the transition sequence of the polyvalent elements (IVB and VB groups) that transform from open-packed and low-symmetry to more compact and symmetrical crystal structures. Thus the $oC16$ structure exists on one side of the Periodic Table in K, Rb and Cs and on the other side in Si, Ge and Bi. Moreover for K and Ge – elements from the same 4$^{th}$ period – this phase occurs with the same lattice parameters in the nearly same pressure range that indicates a possible similarity in their valence electron states [72].

Complex and low-symmetry crystal structures are preferred under compression owing to a formation of many Brillouin planes close to the Fermi surface and the high degree of the BZ filling by electron states which result in significant reduction of the electron band structure energy. Physical properties of these structures are therefore characterized by a significant increase in resistance and decrease in reflectivity. From this point of view it is well expected that at very high compression alkali metals become semiconducting and transparent as was recently found for Li and Na [7,4]. Structural considerations of potassium within the model of overlapping outer core electrons and valence band and satisfying the Hume-Rothery mechanism of stability may be useful for understanding a variety of phenomena in materials under high pressure.

**Acknowledgments**

This work was supported in part by the Programme of the Russian Academy of Sciences "The Matter under High Energy Density".

**References**

[1] McMahon M I and Nelmes R J 2006 High-pressure structures and phase transformations in elemental metals *Chem. Soc. Rev.* **35** 943-63

[2] Degtyareva V F 2006 Simple metals at high pressures: The Fermi sphere - Brillouin zone interaction model *Physics-Uspekhi* **49** 369-88

[3] Degtyareva O 2010 Crystal structure of simple metals at high pressures *High Press. Res.* **30** 343–71

[4] Ma Y, Eremets M, Oganov A R, Xie Y, Trojan I, Medvedev S, Lyakhov A O, Valle M and Prakapenka V 2009 Transparent dense sodium *Nature* **458** 182-5

[5] Shimizu K, Ishikawa H, Takao D, Yagi T and Amaya K 2002 Superconductivity in compressed lithium at 20 K *Nature* **419** 597-9
Struzhkin V V, Eremets M I, GanW, Mao H K and Hemley R J 2002 Superconductivity in dense lithium. *Science* **298** 1213-5
Deemyad S and Schilling J S 2003 Superconducting phase diagram of Li metal in nearly hydrostatic pressures up to 67 GPa *Phys. Rev. Lett.* **91** 167001

[6] Guillaume C L, Gregoryanz E, Degtyareva O, McMahon M I, Hanfland M, Evans S, Guthrie M, Sinogeikin S V and Mao H-K 2011 Cold melting and solid structures of dense lithium *Nature Physics* **7** 211-4

[7] Matsuoka T and Shimizu K 2009 Direct observation of a pressure-induced metal-to-semiconductor transition in lithium *Nature* **458** 186-9

[8] Lundegaard L F, Marqués M, Stinton G, Ackland G J, Nelmes R J and McMahon M I 2009 Observation of the $oP8$ crystal structure in potassium at high pressure *Phys. Rev.* B **80** 020101

[9] Marqués M, Ackland G J, Lundegaard L F, Stinton G, Nelmes R J and McMahon M I 2009 Potassium under pressure: A pseudobinary ionic compound *Phys. Rev. Lett.* **103** 115501

[10] Lundegaard L F, Stinton G W, Zelazny M, Guillaume C L, Proctor J E, Loa I, Gregoryanz E, Nelmes R J and McMahon M I 2009 Observation of a reentrant phase transition in incommensurate potassium *Phys. Rev.* B **88** 054106

[11] Dewaele A, Loubeyre P, and Mezouar M 2004 Equations of state of six metals above 94 GPa *Phys. Rev.* B **70** 094112

[12] Errandonea D, Boehler R, Japel S, Mezouar M, and Benedetti L R 2006 Structural transformation of compressed solid Ar: An x-ray diffraction study to 114 GPa *Phys. Rev.* B **73** 092106

[13] Takemura K, Schwarz U, Syassen K, Hanfland M, Christensen N E, Novikov D L, and Loa I 2000 High-pressure Cmca and hcp phases of germanium *Phys. Rev.* B **62** 10603-6

[14] Chen X J, Zhang C, Meng Y, Zhang R Q, Lin H Q, Struzhkin V V, and Mao H K 2011 *β-tin→Imma→sh* Phase Transitions of Germanium *Phys. Rev. Lett.* **106** 135502 (1-4)

[15] Gregoryanz E, Lundegaard L F, McMahon M I, Guillaume C, Nelmes R J and Mezouar M 2008 Structural diversity of sodium *Science* **320** 1054-57

[16] Ross M and McMahan A K 1982 Systematics of the $s \rightarrow d$ and $p \rightarrow d$ electronic transition at high pressure for the elements I through La *Phys. Rev.* B **26** 4088-93

[17] Degtyareva V F and Degtyareva O 2009 Structure stability in the simple element sodium under pressure *New J. Phys.* **11** 063037

[18] Mott N F and Jones H 1936 *The Theory of the Properties of Metals and Alloys* (London:Oxford University Press)

[19] Jones H 1962 *The Theory of Brillouin Zones and Electron States in Crystals* (Amsterdam: North-Holland)

[20] Hume-Rothery W 1962 *Atomic Theory for Students of Metallurgy* (London: Institute of Metals)

[21] Berger R F, Walters P L, Lee S and Hoffmann R 2011 Connecting the chemical and physical viewpoints of what determines structure: from 1-D chains to gamma-brasses *Chemical Reviews* **111** 4522-45

[22] Degtyareva V F 2003 Brillouin zone concept and crystal structures of sp metals under high pressure *High Press. Res.* **23** 253-57

[23] Feng J, Hoffmann R and Ashcroft N W 2010 Double-diamond NaAl via pressure: Understanding structure through Jones zone activation *J.Chem.Phys.* **132** 14106

[24] Mizutani U 2011 *Hume-Rothery rules for structurally complex alloy phases* (CRC Press, Taylor & Francis, London)

[25] Degtyareva V F and Smirnova I S 2007 BRIZ: a vizualization program for Brillouin zone – Fermi sphere configuration *Z. Kristallogr.* **222** 718-21

[26] Sato H and Toth R S 1962 Fermi Surface of Alloys *Phys. Rev. Lett.* **8** 239-41

[27] King H W 1986 Crystal structures and lattice parameters of allotropes of the metallic elements *Binary Alloys Phase Diagrams* ed. T B Massalski *et al* vol. 2 (Ohio: American Society for Metals, Metals Park) pp 2179-81

[28] Olijnyk H and Holzapfel W B 1983 Phase transitions in K and Rb under pressure *Phys. Lett.* A**99** 381-3

[29] Harrison W A 1966 *Pseudopotentials in the theory of metals* (New York: Benjamin)

[30] McMahon M I, Nelmes R J, Schwarz U and Syassen K 2006 Composite incommensurate K-III and a commensurate form: Study of a high-pressure phase of potassium *Phys. Rev.* B **74** 140102

[31] Hanfland M, Loa I, Syassen K 2002 Sodium under pressure: bcc to fcc structural transition and pressure-volume relation to 100 GPa *Phys. Rev. B* **65** 184109

[32] Hanfland M, Loa I, Syassen K, Schwarz U, Takemura K 1999 Equation of state of lithium to 21 GPa *Solid State Commun.* **112** 123-7

[33] Xie Y, Ma Y M, Cui T, Li Y, Qiu J and Zou G T 2008 Origin of bcc to fcc phase transition under pressure in alkali metals *New J. Phys.* **10** 063022

[34] Dubrovinsky L, Dubrovinskaia N, Crichton W A, Mikhaylushkin A S, Simak S I, Abrikosov I A, deAlmeida J S, Ahuja R, Luo W and Johansson B 2007 Noblest of all metals is structurally unstable at high pressure, *Phys. Rev. Lett.* **98** 045503

[35] Ashcroft N W and Mermin N D *Solid State Physics* 1976 (Brooks/Cole, Cengage Learning) pp 284-92

[36] Hanfland M, Syassen K, ChristensenN E and D.L. Novikov D L 2000 New high-pressure phases of lithium *Nature* **408** 174-8

[37] Degtyareva O, Degtyareva V F, Porsch F and Holzapfel W B 2001 Face-centered cubic to tetragonal transitions in In alloys under high pressure *J. Phys.: Condens. Matter* **13** 7295-303

[38] Kittel C 1995 *Introduction to Solid State Physics* (JohnWiley & Sons, Canada)

[39] Shannon R D 1976 Revised Effective Ionic Radii and Systematic Studies of Interatomic Distances in Halides and Chaleogenides *Acta Cryst.* A**32** 751-67

[40] Villars P (ed) 2002 Pauling File Binaries Edition *Inorganic Materials Database* (Materials Park: ASM International) Software for Materials Science and Engineering

[41] Wyckoff R 1963 *Crystal Structures* (New York: Interscience)

[42] Pearson W B 1972 *The Crystal Chemistry and Physics of Metals and Alloys* (New York: Wiley), pp 454-6

[43] Tremel W, Hoffmann R, Silvestre J 1986 Transitions between NiAs and MnP type phases - an electronically driven distortion of triangular ($3^6$) nets *J. Am. Chem. Soc.* **108** 5174-87.
[44] Ono S, Oganov A R, Brodholt J P, Vočadlo L, Wood I G, Lyakhov A, Glass C W, Côté A S, Price G D 2008 High-pressure phase transformations of FeS: Novel phases at conditions of planetary cores *Earth and Planetary Sci. Lett.* **272** 481-7
[45] Lidin S 1998 Superstructure ordering of intermetallics: B8 structures in the pseudo-cubic regime *Acta Cryst.* B **54** 97-108
Lidin S and Larsson A-K 1995 A survey of superstructures in intermetallic NiAs-$Ni_2$In type phases *J. Solid State Chem.* **118** 313-22
[46] Degtyareva V F, Afonikova N S 2013 Simple metal binary phases based on the body centered cubic structure: Electronic origin of distortions and superlattices *J. Phys. Chem. Solids* **74** 18-24
[47] Von Schnering H G and Nesper R 1987 How Nature Adapts Chemical Structures to Curved Surfaces *Angew. Chem.* **26** 1059-80
[48] Syassen K 2002 Simple Metals at High Pressures *High Pressure Phenomena: Proceedings of the International School of Physics* (Varenna, Italy, 3-13 July 2001) ed R J Hemley and G L Chiarotti (Amsterdam: IOS Press) pp 251-73
[49] Rousseau B and Ashcroft N W 2008 Interstitial Electronic Localization *Phys. Rev. Lett* **101**, 046407
[50] Takemura K, Minomura S and Shimomura O 1982 X-ray diffraction study of electronic transitions in cesium under high pressure *Phys. Rev. Lett.* **49** 1772-5
[51] Schwarz U, Takemura K, Hanfland M and Syassen K 1998 Crystal structure of cesium-V *Phys. Rev. Lett.* **81** 2711-4
[52] Schwarz U, Syassen K, Grzechnik A, and Hanfland M 1999 The crystal structure of rubidium-VI near 50GPa *Solid State Commun.* **112** 319-22
[53] Hanfland M, Schwarz U, Syassen K and Takemura K 1999 Crystal structure of the high-pressure phase silicon VI *Phys. Rev. Lett.* **82** 1197-200
[54] [Chaimayo2012] Chaimayo W, Lundegaard L F, Loa I, Stinton G W, McMahon M I 2012 High-pressure, high-temperature single-crystal study of Bi-IV *High Press. Res.* **32** 442-9
[55] Degtyareva V F, Degtyareva O and Allan D R 2003 Ordered Si-VI-type crystal structure in BiSn alloy under high pressure *Phys. Rev.* B **67** 212105
[56] Degtyareva V F 2000 Crystal structure of a high-pressure phase in Bi-based alloys related to Si VI, *Phys. Rev.* B **62** 9–12
[57] [Schwarz 2000] Schwarz U, Jepsen O, and Syassen K 2000 Electronic structure and bonding in the Cmca phases of Si and Cs *Solid State Commun.* **113** 643-8
[58] Lundegaard L F, Gregoryanz E, McMahon M I, Guillaume C, Loa I and Nelmes R J 2009 Single-crystal studies of incommensurate Na to 1.5 Mbar *Phys. Rev.* B **79** 064105
[59] McMahon M I, Rekhi S and Nelmes R J 2001 Pressure dependent incommensuration in Rb-IV *Phys. Rev. Lett.* **87** 055501
[60] Sakata M, Nakamoto Y, Shimizu K, Matsuoka T and Ohishi Y 2011 Superconducting state of Ca-VII below a critical temperature of 29 K at a pressure of 216 GPa *Phys. Rev.* B **83** 220512
Fujihisa H, Nakamoto Y, Sakata M, Shimizu K, Matsuoka T, Ohishi Y, Yamawaki H, Takeya S, and Gotoh Y 2013 Ca-VII: A chain ordered host-guest structure of calcium above 210 GPa *Phys. Rev. Lett.***110** (accepted)
[61] Alasafi K M and Schubert K 1979 Kristallstrukturen von $Au_3Cd_5$ and $Au_5Cd_8$ h *J. Less-Common Metals* **65** 23-8
[62] McMahon M I, Degtyareva O and Nelmes R J 2000 Ba-IV-type incommensurate crystal structure in group-V metals *Phys. Rev. Lett.* **85** 4896-9
[63] Pickard C J and Needs R J 2010 Aluminium at terapascal pressures *Nature Materials* **9** 624 - 7
Pickard C J and Needs R J 2011 *Ab initio* random structure searching *J. Phys.: Condens. Matter* **23** 053201
[64] Falconi S, McMahon M I, Lundegaard L F, Hejny C, Nelmes R J, and Hanfland M 2006 X-ray diffraction study of diffuse scattering in incommensurate rubidium-IV *Phys. Rev.* B **73** 214102
[65] Degtyareva V F 2010 Electronic origin of the incommensurate modulation in the structure of phosphorus IV *Journal of Physics: Conf. Ser.* **226** 012019
[66] Stager R A and Drickamer H G 1963 Effect of Temperature and Pressure on the Resistance of Four Alkali Metals *Phys. Rev.* **132** 124-7

[67] Takemura K and Syassen K 1983 High-pressure phase transitions in potassium and phase relations among heavy alkali metals *Phys. Rev.* B **28** 1193-6
[68] Stocks G M and Young W H 1969 Electrical resistivities of the alkali metals to pressures of 600 kbar *Phys. C: Solid State Phys.* **2** 680-93
[69] Takemura K, Syassen K 1982 High pressure equation of state of rubidium *Solid State Commun.* **44** 1161-4
[70] Wittig J 1984 *High Pressure in Science and Technology: Proc. of the IX AIRAPT Intern. Conf., Albany, New York, USA, 24 - 29 July 1983 (Materials Research Soc. Symp. Proc.,* vol 22) eds C Homan, R K MacCrone, E Whalley (New York: North-Holland) p. 17
[71] Atou T, Hasegawa M, Parker L J and Badding J V 1996 Unusual chemical behaviour for potassium under pressure: potassium-silver compounds *J. Am. Chem. Soc.* **118** 12104-8
[72] Degtyareva V F 2013 Electronic Origin of the Orthorhombic *Cmca* Structure in Compressed Elements and Binary Alloys *Crystals* **3** 419-30